\documentclass[ aip, amsmath,amssymb,reprint,]{revtex4-1}
\usepackage{graphicx}% Include figure files
\usepackage{dcolumn}% Align table columns on decimal point
\usepackage{bm}% bold math
\usepackage{caption}
\usepackage{comment}
\usepackage{subcaption}
\usepackage{xcolor}
\usepackage{physics}
\usepackage{ulem}
\usepackage{placeins}

 \usepackage{algpseudocode}

\usepackage{algorithm}

\begin{document}

%\title{Examining Nuances of Mixed Finite Element
%Particle-in-Cell with Newmark-Beta Time Stepping \shankernote{Charge conserving mixed finite element PIC that is agnostic to time stepping method}}
\title{Time Integrator Agnostic Charge Conserving Finite Element PIC}
\author{ Scott O'Connor}
 \email{oconn220@msu.edu}
\altaffiliation[Also at ]{Department of Computational Science, Mathematics, and Engineering, Michigan State University, East Lansing, MI}
\author{Zane D. Crawford}
\altaffiliation[Also at ]{Department of Computational Science, Mathematics, and Engineering, Michigan State University, East Lansing, MI}
\author{O. H. Ramachandran}
\altaffiliation[Also at ]{Department of Computational Science, Mathematics, and Engineering, Michigan State University, East Lansing, MI}
\author{John Luginsland}
\affiliation{Department of Electrical and Computer Engineering, Michigan State University, East Lansing, MI}
\author{ B. Shanker}
\affiliation{Department of Electrical and Computer Engineering, Michigan State University, East Lansing, MI}

\begin{abstract}
%Rationale (what exits, what is missing, how to fill the gap)
Developing particle-in-cell (PIC) methods using finite element basis sets, and without auxiliary divergence cleaning methods,  was a long standing problem until recently. It was shown that if consistent spatial basis functions are used, one can indeed create a methodology that was charge conserving, \emph{albeit} using a leap-frog time stepping method. While this is a significant advance, leap frog schemes are \emph{only} conditionally stable and time step sizes are closely tied to the underlying mesh. Ideally, to take full advantage of advances in finite element methods (FEMs), one needs a charge conserving PIC methodology that is \emph{agnostic} to the time stepping method. This is the principal contribution of this paper. In what follows, we shall develop this methodology, prove that both charge and Gauss' laws are discretely satisfied at every time step, provide the necessary details to implement this methodology for both the wave equation FEM and Maxwell Solver FEM, and finally demonstrate its efficacy on a suite of test problems. 
The method will be demonstrated by single particle evolution, non-neutral beams with space-charge, and adiabatic expansion of a neutral plasma, where the debye length has been resolved, and real mass ratios are used.
\end{abstract}

\maketitle

\section{Introduction}\label{sec:introduction}

Simulation of space charge and plasmas is critical to a number of areas in science and engineering. These range from, applications of pulsed power to particle accelerators to  satellites and medicine \cite{marchand2011ptetra,lemke1999three,fourkal2002particle}. The means to do so has largely relied on Particle-in-cell (PIC) methods. PIC has been around since the 1950's and is a popular methods of modeling plasma and space charge due to its simplicity and ease of use \cite{birdsall2004plasma}.  PIC enables a self consistent solution to Maxwell's equation and equations of motion for charged species. Traditionally, PIC is based on finite difference time domain to evolve fields \cite{verboncoeur2005particle}.  The use of regular cubical grids presents challenges, especially in modeling complex geometry. Modeling curved features requires small cell sizes, and this results in a stair-stepped approximation of the desired geometry as well as small time steps in keeping with the Courant–Friedrichs–Lewy condition. Using cut-cells has improved the geometry representation by allowing boundaries to cut across cells \cite{nieter2009application}.
Complex and fine features, as well as multi-scale objects, require the use of a prohibitively expensive number of small cells for high fidelity simulations. As a result of these challenges, there has been persistent investigation into the use of more sophisticated field evolution techniques  \cite{squire2012geometric,monk2003finite,glasser2019geometric,meierbachtol2015conformal}. A natural choice is using time domain finite-element method (TDFEM) due to (a) unconditionally stable time stepping methods, (b) ability to model complex geometries, and (c) well developed extensions to higher order (both in representation of fields and geometry) \cite{jin2015finite}.

While TDFEM can be thought of as a panacea for modeling complex geometries, it is not so for crucial quantities that must be conserved. These include Gauss' law and charge conservation. Indeed, developing a numerical scheme that implicitly conserved charge was an unsolved problem until \cite{pinto2014charge,moon2014exact}. Prior to this development, one used divergence cleaning methods to remove spurious charge accumulation \cite{munz2000divergence}. The key to realizing charge conservation relied on (a) following the \emph{de-Rham} sequence to represent physical quantities on a mesh and (b) use explicit time stepping methods. A more recent paper prescribes three conditions must be satisfied by self-consistent charge conserving schemes \cite{crawford2021rubrics}; this assertion is proved and illustrated for different PIC schemes. The TDFEM-PIC method  relies on Maxwell solvers, in that one solves Maxwell's first order equations as opposed to the wave equation, and leap-frog time stepping. The structure of the solver is such that one avoids a time growing null space corresponding to DC modes. Unfortunately, leap frog is only conditionally stable. As a result, there is a limit on the time-step sizes that one can take, and this closely tied to the underlying discretization. In classical TDFEM, this has been overcome using Newmark-beta time stepping, which is second order and \emph{unconditionally} stable. Unfortunately, implicit time stepping poses a number of challenges to satisfaction of conservation laws that must be satisfied  
and is an open problem\cite{chen2011energy}. This paper provides the theoretical framework for resolving this bottleneck. 

Implicit time stepping permits taking significantly larger time steps, un-constrained by the mesh; and unconditional stability is an added bonus. Unfortunately, as will be evident in the paper, applying these directly to TDFEM-PIC violates both Gauss' law and the equation of continuity. In addition, in solving the field equations, one needs to evolve the locations of particles over time via Newton's laws. A larger time step size, implies that additional infrastructure needs to be in place to accurately compute all aspects of particle trajectory (including information necessary to map it back on the mesh). Resolution to these challenges associated implicit time stepping with a TDFEM framework will be the main contribution of this paper. We will 
\begin{enumerate}
    \item Develop the methods to ensure that both Gauss' law and equation of continuity is satisfied for implicit methods. The methods rely on insight provided in Ref. \cite{crawford2021rubrics}.
    
    \item We will show that the proposed method is agnostic to time stepping schemes. 
    \item We will develop methods to evolve particle parameters (path, velocity along the path, and mapping path to the mesh). 
    
    \item Finally, we will present results validating these methods for \emph{both} the Maxwell and wave equation TDFEM solvers.  
\end{enumerate}
Our hope is to present the technique with sufficient lucidity such that they can be retrofitted with existing codes.

The rest of this paper is organized as follows: In the next Section, we present an overall rubric of implicit TDFEM solvers (both Maxwell and wave), and why direct application of implicit time stepping fails to conserve quantities. Next, in Section \ref{sec:modTDFEM-PIC}, we present details on how these may be modified so as to conserve charge, satisfy Gauss' law, and be independent of time stepping approach. In addition, we present details of the method used to evolve particle parameters. In Section \ref{sec:results}, we present a number of results that validate our claims. Finally, we conclude this paper in Section \ref{sec:conclusions} outlining future directions of research.

\section{Preliminaries \label{sec:Prelim}}

Consider a domain $\Omega$ whose boundaries are denoted by $\partial \Omega$. It is assumed that the domain comprise  charged species that exist in a background medium defined by $\varepsilon_0$ and $\mu_0$, the permittivity and permeability of free space, and the speed of light denoted using $c = 1/\sqrt{\mu_0\varepsilon_0}$; for simplicity of the exposition, we consider only one species. It is also assumed that there exists an electromagnetic field, both impressed and arising from motion of the charged species. Both the fields and the charged species evolve in time.  The distribution of charge can be represented by a phase space distribution function (PSDF) $f(t,\vb{r},\vb{v})$ that satisfies the Vlasov equation 
\begin{align} \label{eq:vlasov}
  \partial_t f(t,\vb{r},\vb{v})  + \vb{v} \cdot \nabla f(t,\vb{r},\vb{v}) + \\ \frac{q}{m} [\vb{E}(t,\vb{r}) + \vb{v} \times \vb{B}(t,\vb{r})] \cdot \nabla_v f(t,\vb{r},\vb{v}) = 0. \nonumber
\end{align}
While we do not solve this equation directly, our approach is conventional in that we make a particle approximation for the PSDF in \eqref{eq:vlasov}. 

\subsection{Overview of Method}

Using this PSDF, we follow the conventional definition of the charge and current density defined as $\rho(t,\vb{r})=q\int_{\Omega} f(t,\vb{r},\vb{v})d\vb{v}$ and $\vb{J}(t,\vb{r})=q\int _{\Omega}\vb{v}(t)f(t,\vb{r},\vb{v})d\vb{v}$ as moments of the PSDF.
 The fields, $\vb{E}(t,\vb{r})$ and $\vb{B}(t,\vb{r})$, in the Vaslov equation are solutions to Maxwell's curl equations with the sources (charge and currents) defined earlier 
\begin{subequations}\label{eq:maxwell_cont}
\begin{equation}\label{eq:faraday}
        - \frac{\partial \vb{B}(t,\vb{r})}{\partial t} = \curl \vb{E}(t,\vb{r})
\end{equation}
\begin{equation}\label{eq:ampere}
      \frac{\partial \vb{D}(t,\vb{r})}{\partial t} = \curl \vb{H}(t,\vb{r}) - \vb{J}(t,\vb{r})
\end{equation}
\end{subequations}
and boundary conditions. These can be either Dirichlet or impedance boundary conditions on $\partial\Omega_D$ or $\partial\Omega_I$, to bound the domain,
\begin{subequations}\label{eq:bceq}
\begin{equation}\label{eq:dirichlet}
\hat{n}\times \mathbf{E}(\mathbf{r},t) = \mathbf{\Psi}_D(\mathbf{r},t)\;\;\text{on}\;\partial\Omega_D,
\end{equation}
\begin{equation}\label{eq:impedanceBC}
\hat{n}\times \frac{\mathbf{B}(\mathbf{r},t)}{\mu} - Y\hat{n}\times\hat{n}\times \mathbf{E}(\mathbf{r},t) = \mathbf{\Psi}_I(\mathbf{r},t)\;\;\text{on}\;\partial\Omega_I.
\end{equation}
\end{subequations}
Instead of using \eqref{eq:maxwell_cont},
the wave equation
 \begin{align}
     \nabla \times \bigg( \frac{1}{\mu_r} \nabla \times  \vb{E} \bigg) + \frac{1}{c_0^2}\epsilon_r  \frac{\partial^2 \vb{E}}{\partial t^2} = -\mu_0 \frac{\partial \vb{J}}{\partial t}
 \end{align}
can be used instead. 
 The magnetic field can be obtained from \eqref{eq:faraday} and the impedance boundary condition is defined using a time derivative on \eqref{eq:impedanceBC} and using \eqref{eq:faraday}.
The fields should also satisfy Gauss' laws
\begin{equation}
    \div \vb{D}(t,\vb{r}) = \rho(t,\vb{r})
\end{equation}
\begin{equation}
    \div \vb{B}(t,\vb{r}) = 0
\end{equation}
though they are not explicitly solved.

As alluded to earlier, we use the moments of PSDF  to find the fields generated and then evolve their position using Newton's equations and Lorentz force, viz., $\vb{F} (t, \vb{r})  = q (t, \vb{r}) (\vb{E} (t, \vb{r}) + \vb{v} (t, \vb{r}) \times \vb{B}(t, \vb{r}) )$, and so on, for the duration of the simulation. Thus far, our description has been in continuous world. To perform an actual simulation, we would need to represent all the quantities involved in terms of  functions defined on a discretization of space and time.  This is typically referred to as a particle in cell (PIC) approach and is the subject of our next discussion. 

Our starting point is the representation of both $\Omega$ and $\partial \Omega$ in terms of a finite set of tetrahedra or a mesh that contains $N_s$ nodes, $N_e$ edges and $N_f$ faces. On these tetrahedra, we define basis functions that follow the \emph{de-Rham} sequence, enabling us to represent fields, fluxes and sources \cite{}. But before proceeding too far ahead, note that we are going to follow the usual PIC cycle; (a) map charges and currents on the mesh, (b) solve for electric and magnetic fields on the mesh, (c) move particles due to Lorentz force and find the current due to this motion, and (d) find the fields due the updated sources. The cycle then continues. 

The starting point of the simulation is to define the charge and currents due to PSDF. With no loss of generality, we follow the usual procedure such that $\rho(t,\vb{r}) = q_{\alpha}\sum^{N_p}_{p=1} \delta(\vb{r}-\vb{r}_p)$ and $\vb{J}(t,\vb{r}) = q_{\alpha}\sum^{N_p}_{p=1}  \vb{v}_p(t) \delta(\vb{r}-\vb{r}_p)$. This implies that PSDF is sampled with $N_p$ shape functions, each being a delta function. Generalization to other shape functions is possible \cite{crawford2021rubrics} and is agnostic to the crux of this paper.

The electric and magnetic fields are represented using Whitney basis functions\cite{monk2003finite,jin2015finite,pinto2014charge}. Specifically,  the electric fields using Whitney edge basis functions,  $\vb{E}(t,\vb{r}) = \sum_{i=1}^{N_e} e_i(t) \vb{W}^{(1)}_{i}(\vb{r})$. The magnetic flux density is represented using  Whitney face basis function,  $\vb{B}(t,\vb{r}) = \sum_{i=1}^{N_f} b_i(t) \vb{W}^{(2)}_{i}(\vb{r})$. Here,   $N_e$ are the number of edges and $N_f$ are the number of faces in the mesh. Two different approaches can be used to solve Maxwell's equations; (a) either solve them in the coupled form or (b) solve the wave equation for the electric field and then obtain the magnetic field. To set the stage for both these solvers, we introduce the following Hodge matrix operators
\begin{equation}
        [\star_\epsilon]_{i,j} = \langle \vb{W}^{(1)}_i(\vb{r}),\varepsilon\cdot\vb{W}^{(1)}_j(\vb{r}) \rangle
\end{equation}
\begin{equation}
        [\star_{\mu^{-1}}]_{i,j} = \langle \vb{W}^{(2)}_i(\vb{r}),\mu^{-1}\cdot\vb{W}^{(2)}_j(\vb{r})\rangle,
\end{equation}
the surface impedance matrix
\begin{equation}
        [\star_I]_{i,j} = \langle \hat{n}_i\cross\vb{W}^{(1)}_i(\vb{r}),\mu^{-1}\cdot\hat{n}_j\cross\vb{W}^{(1)}_j(\vb{r})\rangle
\end{equation}
and discrete curl operator
\begin{equation}
        [\curl]_{i,j} = \langle \vb{\hat{n}}_i , \curl \vb{W}^{(1)}_j(\vb{r})\rangle.
\end{equation}
These matrices are used to build the semidiscrete Maxwell system
\begin{align}\label{eq:maxwell_semi}
\underbrace{\mqty[ [I]& 0 \\ 0 &[\star_{\epsilon_0}]]}_{\Bar{\Bar{C}}_M}\mqty[\partial_t \bar{B}\\ \partial_t \bar{E} ] + \underbrace{\mqty[0&  [\curl] \\ c^2[\curl]^T [\star_{\mu^{-1}}]& c[\star_I] ]}_{\Bar{\Bar{K}}_M} \mqty[\bar{B} \\\bar{E} ] = \underbrace{\mqty[0\\ -\frac{\bar{J}}{\epsilon} ]}_{\bar{\bar{F}}_M}
\end{align}
where the degree of freedom vectors $\bar{E} = [e_1(t),e_2(t),\dots,e_{N_e}(t)]$, $\bar{B} = [b_1(t), b_2(t),\dots,b_{N_f}(t)]$, and $\bar{J}=[j_1(t),j_2(t),...j_{N_e}(t)]$ with $j_i (t) = \langle \vb{W}_i^{(1)} (\vb{r}), \vb{J} (t, \vb{r}) \rangle $.
For the wave equation, the system becomes
\begin{align}\label{eq:wave_cont}
\underbrace{[\star_{\epsilon_0}]}_{\Bar{\Bar{M}}_W}\partial_t^2 \bar{E}  + \underbrace{c[\star_I] }_{\Bar{\Bar{C}}_W} \partial_t\bar{E} +\underbrace{c^2[\star_S]}_{\bar{\bar{K}}_W}\bar{E} = -\partial_t\bar{J} 
\end{align}
where $[\star_S]=[\curl]^T[\star_{\mu^{-1}}][\curl]$.

\subsection{Unconditionally Stable Time Marching\label{sec:MFEM}}

The mixed finite element system in \eqref{eq:maxwell_cont} is discretized in time using Newmark-Beta, an unconditionally stable time stepping method. This method has been extensively used in for the wave equation \cite{jin2015finite} and examined for the mixed finite element method in \cite{crawford2020unconditionally}, allowing for much larger time step sizes than the traditional leapfrog method.
In this method, the fields in time are represented by three temporal basis functions
\begin{equation}
	N_{n+1-i}(t) = \sum_{\substack{j=0 \\ j\neq i}}^{2}
					\frac{t-t_{n+1-j}}{t_{n+1-i}-t_{n+1-j}}
\end{equation}
 corresponding to $i\in [1,3]$ and weighting function
\begin{equation}\label{eq:newmark_test_function}
	W(t) = 
	\begin{cases}
		\frac{t_n-t}{\Delta t} & t \in \left[t_{n-1},t_{n}\right] \\
		\frac{t-t_n}{\Delta t} & t \in \left[t_{n},t_{n+1}\right] \\
		0	& \textrm{otherwise}
	\end{cases}.
\end{equation}

 This combination of basis function and weighting function creates a non-disappative, unconditionally stable time marching scheme, which can be written as recurrence formula provided in \cite{zienkiewicz1977new}, corresponding to parameters $\gamma=0.5$ and $\beta=0.25$. When applied to \eqref{eq:maxwell_cont}, this becomes
 \begin{equation}
 \begin{split}
     (0.5\bar{\bar{C}}_M +0.25\Delta_t\bar{\bar{K}}_M)\bar{X}^{n+1} &-0.5\Delta_t\bar{\bar{K}}_M\bar{X}^n\\
     +(0.5\bar{\bar{C}}_M +.025\Delta_t\bar{\bar{K}}_M)\bar{X}^{n-1} &+0.25\Delta_t\bar{F}_M^{n+1}\\
     +0.5\Delta_t\bar{F}_M^{n} &+ 0.25\Delta_t\bar{F}_M^{n-1}=0
 \end{split}
 \end{equation}
 where $\bar{X}^i=[\bar{B}^{i,T},\bar{E}^{i,T}]$ and $\bar{F}_M^i=[0,-\epsilon^{-1}\bar{J}^{i,T}]$. Likewise, \eqref{eq:wave_cont} becomes
 \begin{equation}
 \begin{split}
     (\bar{\bar{M}}_W + 0.5\Delta_t\bar{\bar{C}}_W +0.25\Delta_t^2\bar{\bar{K}}_W)&\bar{E}^{n+1}\\ +(-2\bar{\bar{M}}_W-0.5\Delta_t^2\bar{\bar{K}}_W)&\bar{E}^n\\
     (\bar{\bar{M}}_W + 0.5\Delta_t\bar{\bar{C}}_W +.025\Delta_t^2\bar{\bar{K}}_W)&\bar{E}^{n-1}\\
     +0.25\Delta_t^2\bar{F}_W^{n+1} +0.5\Delta_t^2\bar{F}_W^{n} + 0.25\Delta_t^2\bar{F}_W^{n-1}&=0
 \end{split}
 \end{equation}
 
 However, treating the current as written in \eqref{eq:maxwell_cont} will not preserve the necessary link between Ampere's law and Gauss' law needed to create a charge conserving scheme. This is immediately apparent after  applying a discrete divergence operator to semidiscrete Ampere's law. After using the identity that $[\div] [\curl]^T = 0$, this operation yields
    \begin{equation}\label{eq:badNmB}
    \begin{split}
        0.5[\div][\star_{\epsilon_0}]\bar{E}^{n+1} &-  0.5[\div][\star_{\epsilon_0}]\bar{E}^{n-1}=\\
      -0.25\Delta_t[\div]\bar{J}^{n+1} &-0.5\Delta_t[\div]\bar{J}^n-0.25\Delta_t[\div]\bar{J}^{n-1}.
    \end{split}
    \end{equation}
    When the same operator is applied to the semidiscrete wave equation, one gets
    \begin{equation}\label{eq:badwave}
    \begin{split}
        [\div][\star_{\epsilon_0}]\bar{E}^{n+1}&-2[\div][\star_{\epsilon_0}]\bar{E}^n+ [\div][\star_{\epsilon_0}]\bar{E}^{n-1}\\
        -0.25\Delta_t[\div]\bar{J}^{n+1}&-0.5\Delta_t[\div]\bar{J}^n-0.25\Delta_t[\div]\bar{J}^{n-1}.
    \end{split}
    \end{equation}
Making the substitution of $\bar{\rho}^i = [\div][\star_{\epsilon_0}]\bar{E}^i$, then it becomes clear that neither \eqref{eq:badNmB} nor \eqref{eq:badwave} satisfy Gauss' law or the continuity equation. Instead, a different treatment of the right hand side, the particle current density, is needed in order to create a charge conserving scheme. 

\section{Modified TDFEM-PIC \label{sec:modTDFEM-PIC}}

\subsection{Integrator Agnostic Charge Conserving Scheme}

It is apparent that, as written, time conservation fails for both Maxwell solver and the wave equation. The reasons are two fold: (a) the order of time derivatives on the current (on the right hand side) and those on the electric field are off by one; (b) this requires the discrete time integrator to remember initial conditions. The latter holds the key to solving the puzzle. Newmark time stepping schemes are, in effect, stable time integrators.  The crux of our approach is to correctly evaluate the time integral of the current. As elucidated in \cite{crawford2021rubrics}, the time integral of the current is readily obtainable, and indeed a part of the PIC scheme. Specifically, starting with the definition of the PDSF, 
\begin{equation}\label{eq:chargeDef}
       - \bar{\rho}^{n} (\vb{r}(t_n))  = [\div]\int_0^{t_n} \bar{J}^n(\vb{r}(\tau) ) d\tau.
\end{equation}
where $t_n = n\Delta_t$. As shown in \cite{crawford2021rubrics} this equation can be rewritten as, 
\begin{equation}\label{eq:gaussredo}
        \bar{\rho}^{n} (\vb{r}(t_n)) = -  [\div]\int_{\vb{r}(0)}^{\vb{r}(t_n)} \bar{J}^i(\tilde{\vb{r}}) d\tilde{\vb{r}}.
\end{equation}
Following the details in Ref. \cite{crawford2021rubrics}, it is immediately apparent that for any particle $p$
\begin{equation}
 \int_0^{t_n} d\tau  \vb{v}_p(\tau)\delta(\vb{r}- \vb{r}_p(\tau)) =  \int_{\vb{r}_p(0)}^{\vb{r}_p(t_n)} d\tilde{\vb{r}} \delta\left(\vb{r} -\tilde{\vb{r} } \right)
\end{equation}

Note, for each charge, its trajectory is determined by the solution to Newton's equations. The integration along a particle path can be computed to a high degree of accuracy. To develop a charge conserving methodology, we define 
\begin{align}\label{eq:G}
    \bar{G}^n = \int_0^{t_n} \bar{J}^n(\tau) d\tau
\end{align}
This equation is readily evaluated using \eqref{eq:gaussredo}. It follows that instead of using $\bar{J}$ in \eqref{eq:maxwell_cont} (and therefore, in \eqref{eq:badNmB}) one can instead use $\partial_t\bar{G}$. Discrete implementation with a Maxwell equation solver results in in the divergence of Ampere's law to be
    \begin{equation}\label{eq:goodNmB}
    \begin{split}
        0.5[\div][\star_{\epsilon_0}]\bar{E}^{n+1} &-  0.5[\div][\star_{\epsilon_0}]\bar{E}^{n-1}=\\
      -.5[\div]\bar{G}^{n+1}&+0.5[\div]\bar{G}^{n-1}.
    \end{split}
    \end{equation}
    Examining \eqref{eq:goodNmB} term by term reveals that both sides of the equation are identical given that $[\div] \bar{G}^n = - \bar{\rho}^n$. In a similar manner, one can use  $\partial_t^2\bar{G}$ instead of $\partial_t\bar{J}$ in \eqref{eq:wave_cont} to yield
    \begin{equation}\label{eq:goodNmB}
        \bar{\rho}^{n+1} -2\bar{\rho}^{n}+\bar{\rho}^{n-1}=
      -[\div]\bar{G}^{n+1}-2[\div]\bar{G}^{n-1}+[\div]\bar{G}^{n-1}.
    \end{equation}
Here, we have taken the liberty of substituting, $\bar{\rho}^n = [\div][\star_{\epsilon_0}]\bar{E}^n$. At this point, we note that the proposed approach is \emph{agnostic} to the time stepping scheme (or integrator) used to solve Maxwell's equations; both the equation of continuity and Gauss' laws are satisfied by design.

A word of caution is in order before we proceed. While, the method developed is exact, it should be noted that to obtain $\bar{E}^{n+1}$, one needs to solve either \eqref{eq:badNmB} (or \eqref{eq:badwave}) with the appropriate substitutions for $\bar{G}^{n+1}$ instead of $\bar{J}^{n+1}$. Obviously, the solution to these sets of equations is subject to errors that arise due to vagaries of a linear algebraic solution (tolerances, excitation of null-spaces, etc). As a result, as will be seen in the results section, our errors are small but not identically zero. Next, we discuss a higher order particle pusher to solve the equations of motion consistently.
    
\subsection{Particle Pusher}

Using an implicit time stepping scheme has advantages as well as challenges. The principal advantage is an unconditionally stable time step size independent scheme as opposed to a conditionally stable scheme like leap frog whose stability depends on the time step size. The downside of using large time steps is that one must capture the nuances of both the path and velocity of the particle. Thus, solving the equations of motion using a Boris push\cite{boris1970relativistic}, with its linear representation of the particle position and velocity can introduce large errors into the system. Our goal is to develop a higher order scheme.

As is well known, the  particle positions and velocities are updated by solving Newton's 
equations via the Lorentz force, giving us the following set of coupled first 
order ODEs for each particle,
\begin{align}\label{eq:newtons}
    \partial_t \vb{v}_p (t, \vb{r}_p) &= \vb{a}_p (t,\vb{r}_p) =  \frac{q_{\alpha}}{m_{\alpha}}\left (\vb{E}(t,\vb{r}_p ) + \vb{v}_p (t, \vb{r}_p) \times \vb{B}(t,\vb{r}_p )\right ) \\
    \partial_t \vb{r}_p (t, \vb{r}_p) &= \vb{v}_p (t, \vb{r}_p).
\end{align}
These form a pair of first order ODEs in time, and there are a number of methods that can be applied. Our choice is to use a higher order order Adams-Bashforth scheme. An exemplar recursion relation for $\vb{v}$ and $\vb{r}$ for a 4th order Adam's-Bashforth method is as follows:
\begin{align}\label{eq:adam-bashforth}
    \vb{v}_p^{n+1} &= \vb{v}_p^{n} + \frac{\Delta_t}{24}(55 \vb{a}_p^n - 59 \vb{a}_p^{n-1} + 37 \vb{a}_p^{n-2}-9\vb{a}_p^{n-3})\\
    \vb{r}_p^{n+1} &= \vb{r}_p^{n} + \frac{\Delta_t}{24}(55 \vb{v}_p^n - 59 \vb{v}_p^{n-1} + 37 \vb{v}_p^{n-2}-9\vb{v}_p^{n-3}).\label{eq:adam-bashforth-pos}
\end{align}
where $\Delta_t$ is the time step size. Given that the Newmark scheme is second order, we choose the Adams-Bashforth scheme to be at least two orders higher so as to accommodate a second time derivative in on $\bar{G}^n$.
 The path used for interpolating the position is a fourth order Lagrange polynomial $k=4+1$ is defined as,
\begin{subequations}
\label{eq:path}
\begin{align}
    \vb{r}_p(t) &= \sum_{j=0}^k\vb{r}^{n-j}_p \ell_j(t) \\
    \ell(t) &= \prod_{ \begin{subarray}{c}0 \le m \le k \\ m \ne j\end{subarray}} \frac{t - t_{n+1-m}}{t_{n+1-j} - t_{n+1-m}},
\end{align}
\end{subequations}
where $\vb{r}_p(t)$ is the position at time $t$ and $\vb{r}_p^n$ is the location of particle $p$ at the $t=n\Delta_t$. %\shankernote{do no understand}

\subsection{Particle Path and Current Mapping}

The final piece of the puzzle is mapping the path to the underlying tesselation. In order to do so, we note that that the integrator used to solve the equation of motion implicitly assumes a Lagrange polynomial interpolant. As a result, the order of the method used maps to order of the interpolant. This information needs to be used to find out where the particle enters and leaves the cell.

Once the particle locations at each time step are known from the particle push, the path through the unstructured mesh needs to be found. 
This includes finding the locations of where a particle enters a cell and where it leaves and is detailed in Algorithm \ref{alg:particle_path_algorithm}.
Since we are using a higher order representation of a particle path, finding these entry and exit points of the cell with the tetrahedron becomes a non-linear problem and is detail in Algorithm \ref{alg:bisection_method}. Assume that we are given the normal to surface $\hat{\vb{n}}$, and vertices of the triangle $\vb{r}_{v,1}$, $\vb{r}_{v,2}$, $\vb{r}_{v,3}$. The intersection between the trajectory $\vb{r}_p(t)$ and the plane can be obtained by  solving 
\begin{equation}
\label{eq:intersection}
     \hat{\vb{n}}\cdot \left [  \left ( \vb{r}_p(t) - \vb{r}_{v,1} \right ) \times \left ( \vb{r}_{v,2} - \vb{r}_{v,1}\right ) \right ]  = 0  
\end{equation}

 \begin{algorithm}[H]
\caption{Particle Path Finding Algorithm}
\label{alg:particle_path_algorithm}
\begin{algorithmic}[1]
\State Push particle finding $\vb{r}_{p,f}$
\If{$\vb{r}_{p,f}$ is in same cell as $\vb{r}_{p,s}$}
    \If{All quadrature points between are in same cell} 
        \State Integrate using a quadrature rule along path. 
        \State \textbf{Return} and go onto next particle
    \EndIf
\EndIf
\State Find exit point $\vb{r}_{p,i}$ of path in cell
\State Integrate from $\vb{r}_{p,s}$ to $\vb{r}_{p,i}$
\While{Path not complete}
\State Find next cell that path travels through
\If{$\vb{r}_f$ is in same cell as $\vb{r}_{p,i}$}
    \If{All quadrature points are in same cell} 
        \State Integrate using a quadrature rule along path. 
        \State Path is complete
        \State \textbf{Return} and go onto next particle.
    \Else
        \State Find exit point of path in cell.
        \State Integrate from $\vb{r}_{p,is}$ to $\vb{r}_{p,if}$.
    \EndIf
\EndIf
\EndWhile
\end{algorithmic}
\end{algorithm} 
  
 \begin{algorithm}[H]
\caption{Non-Linear Bi-Section Method}
\label{alg:bisection_method}
\begin{algorithmic}[1]
\State $t_s = 0$,$t_f=1$,$t_h=0.5$
\If{Any quadrature points are outside of the cell.}
\State $t_f=t_q$
\EndIf
%\State Push particle finding $\vb{r}_f$
\While{$|t_s-t_f|< tol$}
\If{$\vb{t}_h$ is in same cell as $\vb{t}_s$}
    \State $t_s=t_h$
\Else
    \State $t_f=t_h$   
\EndIf
    \State $t_h = 0.5(t_s+t_f)$
\EndWhile
\end{algorithmic}
\end{algorithm} 

\begin{figure}
    \centering
    \includegraphics{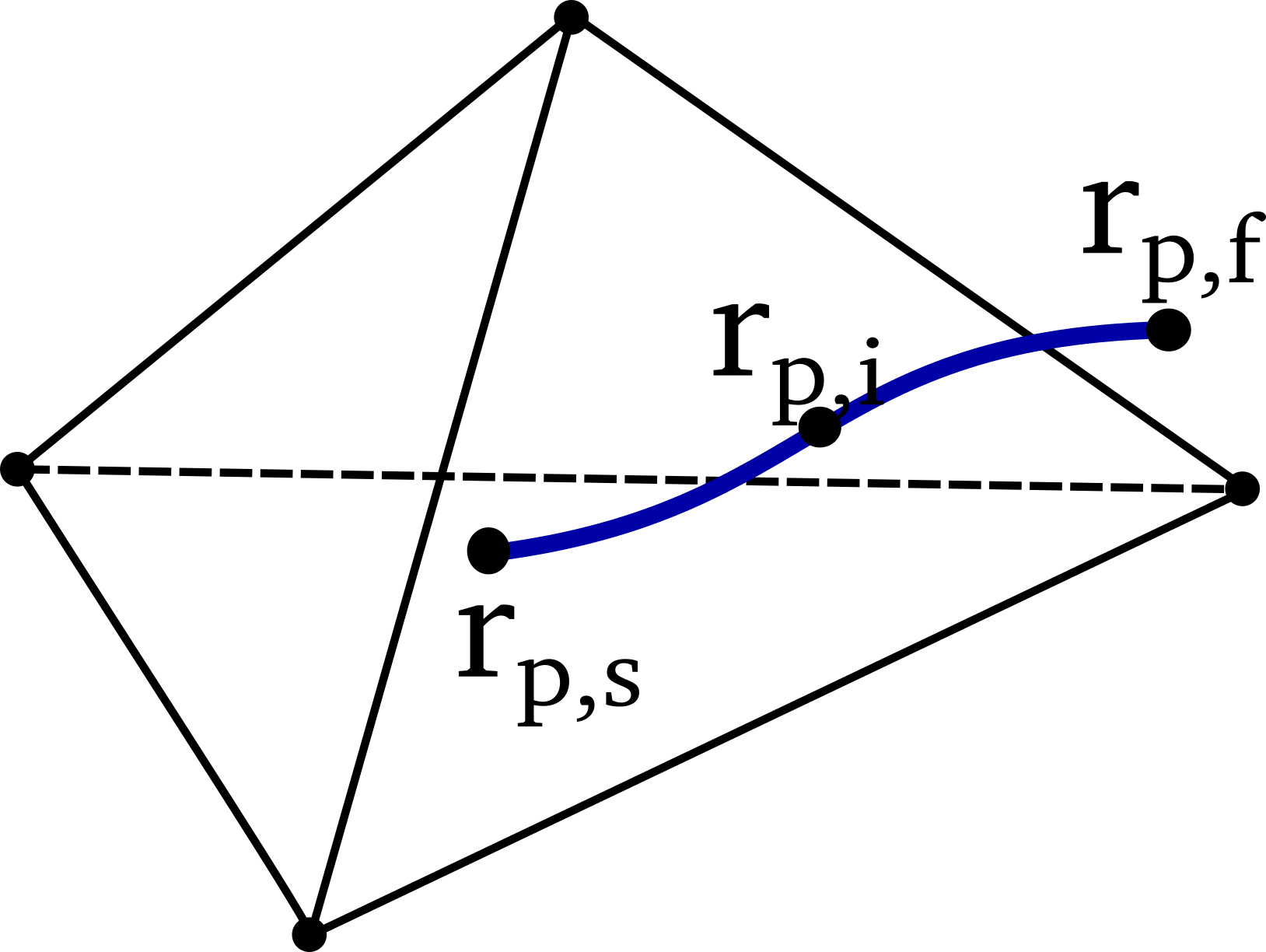}
    \caption{Particle path for a single particles with start and location and intersection point.}
    \label{fig:particle_path}
\end{figure}

Note, the path $\vb{r}_p(t)$ can be parameterized using  \eqref{eq:path}. Using this parameterization, one can use a non-linear iteration (such as Newton-Raphson) to solve \eqref{eq:intersection}. For convinience, we take a simpler approach by implementing a bi-section method that moves along the path checking whether candidate points are  inside or outside of the cell. For test cases presented in this paper, this method converges rather robustly. Every step takes around 47 steps to converge below a tolerance of $1\cdot10^{-15}$ ($0.5^{47}=7.1\cdot10^{-15}$). 
Once the method converges, we then compute the integral along each path segment in each cell using a set of quadrature points. 
To illustrate this, consider Fig. \ref{fig:particle_path} containing an example particle starting position $\vb{r}_{p,s}$, and finishing position $\vb{r}_{p,f}$ with the intersection point being $\vb{r}_{p,i}$.
The quadrature points would lie along the path between the $\vb{r}_{p,s}$ and $\vb{r}_{p,i}$, then another set of quadrature points between $\vb{r}_{p,i}$ and $\vb{r}_{p,f}$. 

Before we discuss results obtained using the above approach, a few points are in order; to evaluate $\bar{G}^n$, (a) the integral over the path be evaluated using quadrature rules to very high precision as the order of the path is know; (b) when the path passes though multiple cells, the integration is broken up into pieces over each cell; (c) one can save on computational cost of by updating the integral.

\section{Results\label{sec:results}}

In this Section, we present a number of results demonstrating the efficacy of the proposed scheme with respect to conservation laws, as well accuracy of key steps that are integral to the process.

\subsection{Higher Order Particle Motion}

One of the key advantages in using implicit time stepping is the possibility of using much larger time step sizes. Unfortunately, this also implies that one needs higher order methods to capture both the path as well as velocity. In this section, we demonstrate convergence of our algorithm for particle motion using various orders of Adams-Bashforth integrator and compare these to standard non-relativistic Boris push. 

\begin{figure}
    \centering
    \includegraphics[scale=0.5]{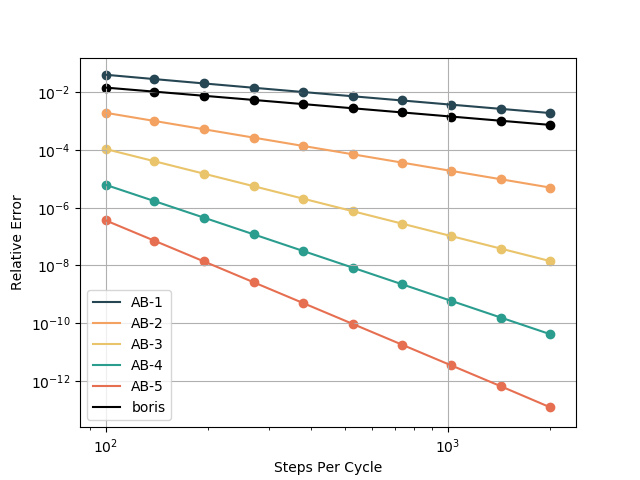}
    \caption{Mean relative error in position for Adam-Bashforth Orders 1-5 compared with the Boris push, shown in black.}
    \label{fig:cyclotron_motion}
\end{figure}

\begin{figure}
    \centering
    \includegraphics[scale=0.5]{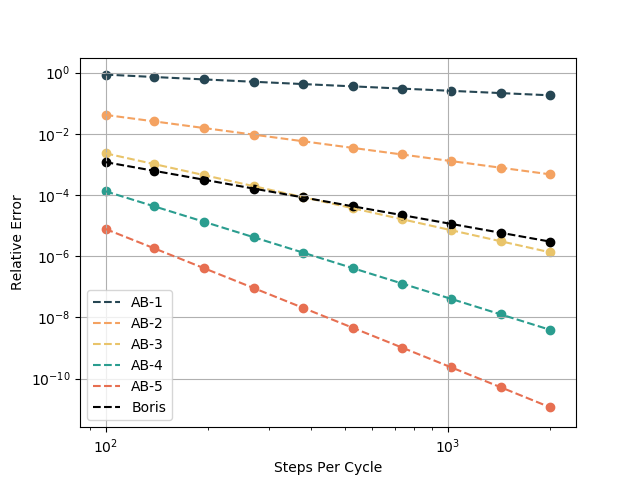}
    \caption{Mean relative error in velocity for Adam-Bashforth Orders 1-5 compared with the Boris push, shown in black.}
    \label{fig:ab_v_boris}
\end{figure}

To do so,  we set up a classic cyclotron \cite{o2021set} motion test where a single particle was given an initial velocity in a constant magnetic field resulting in circular motion as shown in Fig. \ref{fig:ab_v_boris}.   
The parameters are shown in Table \ref{tb:cyclotron} with a particle's initial velocity $\vb{v}_0$ with a background magnetic fields $\vb{B}$ with a given mass $m$ and charge $q$. The particle will move in a circle due to the Lorentz force as shown in Fig. \ref{fig:cyclotron_motion}.
The relative error in both position and velocity for various time step sizes with multiple order of Adams-Bashforth and Boris are is shown in Fig. \ref{fig:ab_v_boris}.
The average error is calculated by taking the norm of the distance errors of each point $\vb{r}$ and dividing by the normal of the analytic positions $\vb{r}_a$ (see Ref. \cite{o2021set} for details),
\begin{equation}
    error = \frac{||\vb{r} - \vb{r}_a||_2}{||\vb{r}_a||_2}
\end{equation}
The slopes for each of the Adams-Bashforth methods match its order. Boris on the other hand has a second order velocity update with a first order positional update. This test essentially validates out pusher as well as helps correlate error (or approximately so) in particle motion with time step size. 

\begin{table}[ht!]
\centering
    \caption{Cyclotron Motion}
    \begin{tabular}{c|c}
         \textit{Parameter} & \textit{Value}  \\
         \hline
         $\vb{B}$ & $6.822756\cdot10^{-5} \hat{z}$ T \\
         Q & $-1.60217646\cdot 10^{-19}$ C  \\
         m & $9.10938370 \cdot 10^{-31}$ kg \\
         $\vb{v}_0$ & $3\cdot 10^6 \hat{y}$ m/s \\ 
         $\vb{r}_0$ & $[0.75,0.5,0.0]$ m
    \end{tabular} \label{tb:cyclotron}
\end{table}

\subsection{Expanding Particle Beam}

Next, we consider an expanding beam test \cite{o2021set}. An expanding particle beam is injected into a cylindrical cavity with an initial velocity of magnitude $v_0$. 
As the beam travels down the tube, the electrons repel each other causing the beam beam to expand. This expansion rate can be compared with other codes to validate the solution. 
The detail of the mesh and beam parameters used are shown in Table \ref{tb:beam}.
\begin{table}[ht!]
\centering
    \caption{Expanding Particle Beam Parameters}
    \begin{tabular}{c|c}
         \textit{Parameter} & \textit{Value}  \\
         \hline
         Cavity Radius & 20 mm \\
         Cavity Length & 100 mm\\
         Boundary Conditions & PEC \\
         $v_0$ & $5\cdot10^7$ m/s \\ 
         $v_0/c$ & 0.16678 \\ 
         beam radius $r_b$ & 8.00 mm \\
         Number particles per time step & 10 \\
         species & electrons \\
         Turn on time & 2 ns \\
         beam current  & 0.25 A \\
         macro-particle size & 52012.58 \\
         min edge length & 1.529 mm \\
         max edge length & 6.872 mm \\
         $\Delta_t$ & ns\\
    \end{tabular} \label{tb:beam}
\end{table}

\begin{figure}
    \centering
    \includegraphics[scale=0.5]{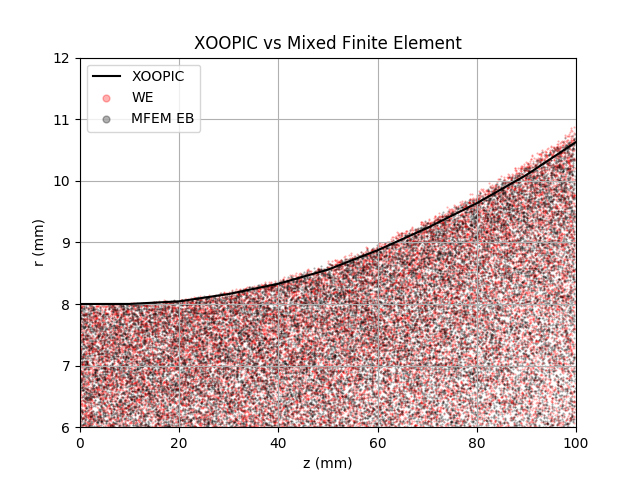}
     \caption{Expanding particle beam macro particles in the z vs r plan. Particle locations from both mixed finite element methods and wave equation versions are compared with XOOPIC beam profile.}
    \label{fig:we_vs_mfem_traj}
\end{figure}

\begin{figure}
    \centering
    \includegraphics[scale=0.5]{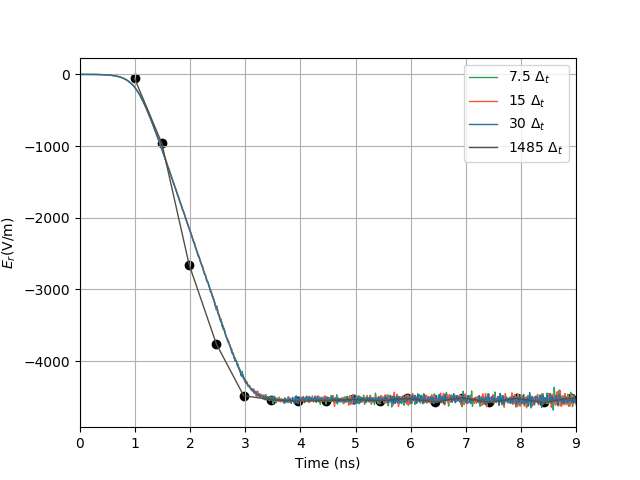}
    \caption{Electric field values are the radial component half way down  the tube 16 mm from the center of the tube. Multiple simulation with different time steps are performed.}
    \label{fig:efield_comparison}
\end{figure}

Both the wave equation and mixed finite element trajectories are compared in Fig. \ref{fig:we_vs_mfem_traj} and show good agreement with XOOPIC \cite{verboncoeur2005particle} (an extensively used and well validated quasi-2D FDTD code). 
We sample the electric field half way down the tube 16 mm from the center of the tube. The radial field values are plotted over time shown in  Fig. (\ref{fig:efield_comparison}) for  simulations with different time steps.  We compare four runs with time steps of $\alpha \Delta_t$ where $\alpha$ is scale factor and $\Delta_t = 0.333$ ps is the largest stable step size in a leap frog time marching method for the given mesh. 
Note, 2 ns corresponds to 1 transit of the tube. 
It is evident from this figure that the proposed method provides stable results; indeed, as is evident from this figure, the data at 7.5$\Delta_t$, $15 \Delta_t$ and $30 \Delta_t$ are almost identical to each other, where as the one at $1485 \Delta_t$ is slightly different. This points to significant gains that can be made with Newmark time stepping (provided the method is charge conserving). 

\begin{figure}
    \centering
    \includegraphics[scale=0.5]{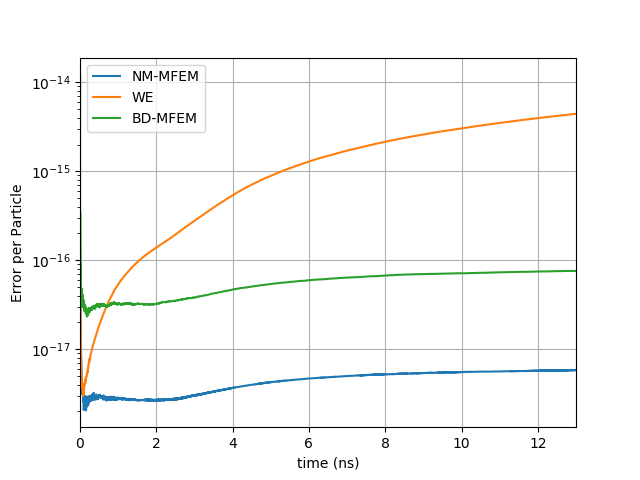}
    \caption{Discrete Gauss's Law error per particle for Newmark-Beta mixed finite element (NM-MFEM), Newmark-Beta wave equation (WE), backwards difference mixed finite elements (BD-MFEM) using the charge conservation technique provided here. }
    \label{fig:DGL_beam}
\end{figure}

This leads to the next argument. Shown in Fig. \ref{fig:DGL_beam} is data from two different methods for the same set up run using MFEM with backward difference at $\Delta_t$, MFEM with Newmark at $7\Delta_t$ and the wave equation (WE) at $7 \Delta_t$. As evident, all three methods conserve charge to almost machine precision. It should be noted that both MFEM and WE have a null space. In the case of the former, it is fields that behave like $\nabla \phi (\vb{r})$, and the latter, as $t \nabla \phi (\vb{r})$. However, as is evident from these results, our mapping on to these null spaces is small and behaves as expected. 

\begin{figure}
    \centering
    \includegraphics[scale=0.5]{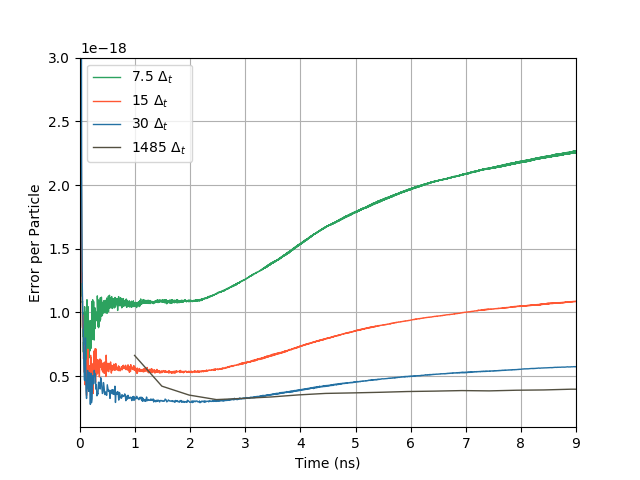}
    \caption{Discrete Gauss's law error per particle various time steps using the mixed finite element methods using Newmark time stepping.  }
    \label{fig:conservLaw_comparison}
\end{figure}

To further illustrate the robustness of the method to time step sizes, in Fig. \ref{fig:conservLaw_comparison} we compare the satisfaction of Gauss' law for all four time steps used in Fig. \ref{fig:efield_comparison}. As is evident from here, charge is again conserved almost to machine precision (around $10^{-18}$ for all with slight difference evolution of trajectory). 

\subsection{Adiabatic Expanding Plasma}

\begin{figure}
    \centering
    \includegraphics[scale=0.5]{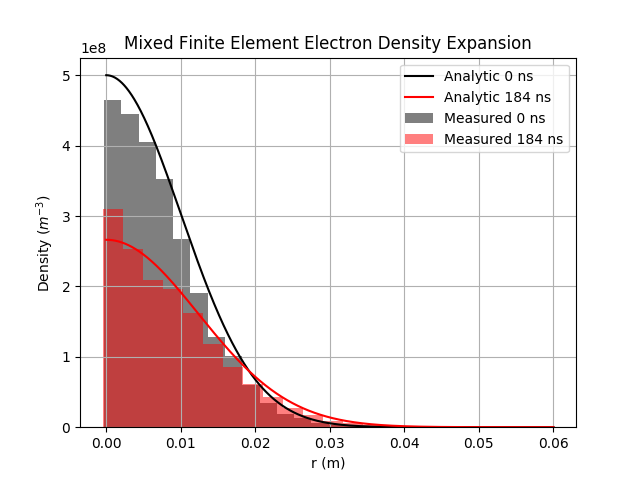}
    \caption{Mixed Finite Element particle beam expansion}
    \label{fig:plasma_ball_MFEM}
\end{figure}

\begin{figure}
    \centering
    \includegraphics[scale=0.5]{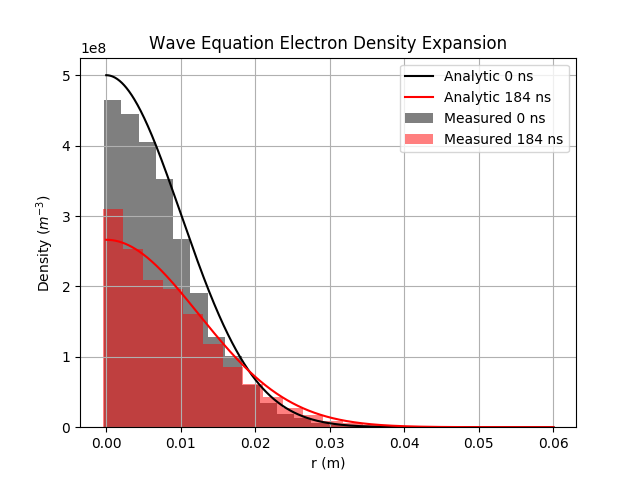}
    \caption{Wave Equation Adiabatic Expanding Plasma}
    \label{fig:plasma_ball_WE}
\end{figure}

Finally, for a third validation case we simulate an adiabatic expansion of a plasma ball with radial Gaussian distribution in the radial direction.
This case has an analytic solutions \cite{kovalev2003analytic} and allows for good comparison and validation. 
We change some of the parameters from the original numerical experiments \cite{o2021set} such that the Debye length can be fully resolved.
This example is described in more detail in\cite{o2021set}.
We simulate this example both  MFEM and WE. For both examples we get excellent agreement in the expansion rate with both the wave equation, Fig. (\ref{fig:plasma_ball_WE}), and the mixed formulation, Fig. \ref{fig:plasma_ball_MFEM}, when compared with analytic densities. 
\begin{table}[]
    \centering
    \caption{Adiabatic Expanding Plasmas}
    \begin{tabular}{c|c} \label{tabe:plasma_ball}
         \textit{Parameter} & \textit{Value}  \\
         \hline
         Mesh Radius & 6mm \\
         Boundary Conditions & First order ABC \\
         $T_{ion}$ & 1K \\ 
         $T_{electron}$ & 100K \\ 
         Number Particles & 8000 \\
         Species & Electrons and $Sr^{+}$ \\
         Macro-Particle Size & 52012.58 \\
         Min Edge Length & 1.529mm \\
         Max Edge Length & 6.872mm \\
    \end{tabular}
\end{table}


\section{Summary \label{sec:conclusions}}

In this paper, we have presented a solution to a problem that has been long-standing--charge conserving FEM-PIC methods for implicit time stepping systems \emph{without} the need to adopt divergence cleaning. In other words, rubrics have been developed such that conservation laws are implicitly obeyed. Indeed, the method presented is agnostic to any time stepping scheme. We have demonstrated the efficacy of this approach for a set of test problems, using different time step sizes and different time stepping schemes, as well as both MFEM and WE solvers. The results reliably attest our claims. The above approach opens multiple doors that will further the state of art of FEM-PIC; these include higher order schemes in both space and time, quasi-Helmholtz decomposition to get a better handle on null-spaces, and domain decomposition to effect rapid solution by parallelizing the scheme. Papers on these will be presented soon in other forums. 

\begin{acknowledgments}
This work was supported by SMART Scholarship program. We thank the MSU Foundation for support through the Strategic Partnership Grant during early portion of this work. This work was also supported by the Department of Energy Computational Science Graduate Fellowship under grant DE-FG02-97ER25308. The authors would also like to thank the HPCC Facility,
Michigan State University, East Lansing, MI, USA.
\end{acknowledgments}

\section*{Data Availability \label{sec:data}}
The data that support the findings of this study are
available from the corresponding author upon reasonable
request.

\bibliography{HOtime}

\begin{thebibliography}{21}%
\makeatletter
\providecommand \@ifxundefined [1]{%
 \@ifx{#1\undefined}
}%
\providecommand \@ifnum [1]{%
 \ifnum #1\expandafter \@firstoftwo
 \else \expandafter \@secondoftwo
 \fi
}%
\providecommand \@ifx [1]{%
 \ifx #1\expandafter \@firstoftwo
 \else \expandafter \@secondoftwo
 \fi
}%
\providecommand \natexlab [1]{#1}%
\providecommand \enquote  [1]{``#1''}%
\providecommand \bibnamefont  [1]{#1}%
\providecommand \bibfnamefont [1]{#1}%
\providecommand \citenamefont [1]{#1}%
\providecommand \href@noop [0]{\@secondoftwo}%
\providecommand \href [0]{\begingroup \@sanitize@url \@href}%
\providecommand \@href[1]{\@@startlink{#1}\@@href}%
\providecommand \@@href[1]{\endgroup#1\@@endlink}%
\providecommand \@sanitize@url [0]{\catcode `\\12\catcode `\$12\catcode
  `\&12\catcode `\#12\catcode `\^12\catcode `\_12\catcode `\%12\relax}%
\providecommand \@@startlink[1]{}%
\providecommand \@@endlink[0]{}%
\providecommand \url  [0]{\begingroup\@sanitize@url \@url }%
\providecommand \@url [1]{\endgroup\@href {#1}{\urlprefix }}%
\providecommand \urlprefix  [0]{URL }%
\providecommand \Eprint [0]{\href }%
\providecommand \doibase [0]{http://dx.doi.org/}%
\providecommand \selectlanguage [0]{\@gobble}%
\providecommand \bibinfo  [0]{\@secondoftwo}%
\providecommand \bibfield  [0]{\@secondoftwo}%
\providecommand \translation [1]{[#1]}%
\providecommand \BibitemOpen [0]{}%
\providecommand \bibitemStop [0]{}%
\providecommand \bibitemNoStop [0]{.\EOS\space}%
\providecommand \EOS [0]{\spacefactor3000\relax}%
\providecommand \BibitemShut  [1]{\csname bibitem#1\endcsname}%
\let\auto@bib@innerbib\@empty
%</preamble>
\bibitem [{\citenamefont {Marchand}(2011)}]{marchand2011ptetra}%
  \BibitemOpen
  \bibfield  {author} {\bibinfo {author} {\bibfnamefont {R.}~\bibnamefont
  {Marchand}},\ }\bibfield  {title} {\enquote {\bibinfo {title} {Ptetra, a tool
  to simulate low orbit satellite--plasma interaction},}\ }\href@noop {}
  {\bibfield  {journal} {\bibinfo  {journal} {IEEE Transactions on Plasma
  Science}\ }\textbf {\bibinfo {volume} {40}},\ \bibinfo {pages} {217--229}
  (\bibinfo {year} {2011})}\BibitemShut {NoStop}%
\bibitem [{\citenamefont {Lemke}, \citenamefont {Genoni},\ and\ \citenamefont
  {Spencer}(1999)}]{lemke1999three}%
  \BibitemOpen
  \bibfield  {author} {\bibinfo {author} {\bibfnamefont {R.}~\bibnamefont
  {Lemke}}, \bibinfo {author} {\bibfnamefont {T.}~\bibnamefont {Genoni}}, \
  and\ \bibinfo {author} {\bibfnamefont {T.}~\bibnamefont {Spencer}},\
  }\bibfield  {title} {\enquote {\bibinfo {title} {Three-dimensional
  particle-in-cell simulation study of a relativistic magnetron},}\ }\href@noop
  {} {\bibfield  {journal} {\bibinfo  {journal} {Physics of Plasmas}\ }\textbf
  {\bibinfo {volume} {6}},\ \bibinfo {pages} {603--613} (\bibinfo {year}
  {1999})}\BibitemShut {NoStop}%
\bibitem [{\citenamefont {Fourkal}\ \emph {et~al.}(2002)\citenamefont
  {Fourkal}, \citenamefont {Shahine}, \citenamefont {Ding}, \citenamefont {Li},
  \citenamefont {Tajima},\ and\ \citenamefont {Ma}}]{fourkal2002particle}%
  \BibitemOpen
  \bibfield  {author} {\bibinfo {author} {\bibfnamefont {E.}~\bibnamefont
  {Fourkal}}, \bibinfo {author} {\bibfnamefont {B.}~\bibnamefont {Shahine}},
  \bibinfo {author} {\bibfnamefont {M.}~\bibnamefont {Ding}}, \bibinfo {author}
  {\bibfnamefont {J.}~\bibnamefont {Li}}, \bibinfo {author} {\bibfnamefont
  {T.}~\bibnamefont {Tajima}}, \ and\ \bibinfo {author} {\bibfnamefont {C.-M.}\
  \bibnamefont {Ma}},\ }\bibfield  {title} {\enquote {\bibinfo {title}
  {Particle in cell simulation of laser-accelerated proton beams for radiation
  therapy},}\ }\href@noop {} {\bibfield  {journal} {\bibinfo  {journal}
  {Medical Physics}\ }\textbf {\bibinfo {volume} {29}},\ \bibinfo {pages}
  {2788--2798} (\bibinfo {year} {2002})}\BibitemShut {NoStop}%
\bibitem [{\citenamefont {Birdsall}\ and\ \citenamefont
  {Langdon}(2004)}]{birdsall2004plasma}%
  \BibitemOpen
  \bibfield  {author} {\bibinfo {author} {\bibfnamefont {C.~K.}\ \bibnamefont
  {Birdsall}}\ and\ \bibinfo {author} {\bibfnamefont {A.~B.}\ \bibnamefont
  {Langdon}},\ }\href@noop {} {\emph {\bibinfo {title} {Plasma physics via
  computer simulation}}}\ (\bibinfo  {publisher} {CRC press},\ \bibinfo {year}
  {2004})\BibitemShut {NoStop}%
\bibitem [{\citenamefont {Verboncoeur}(2005)}]{verboncoeur2005particle}%
  \BibitemOpen
  \bibfield  {author} {\bibinfo {author} {\bibfnamefont {J.~P.}\ \bibnamefont
  {Verboncoeur}},\ }\bibfield  {title} {\enquote {\bibinfo {title} {Particle
  simulation of plasmas: review and advances},}\ }\href@noop {} {\bibfield
  {journal} {\bibinfo  {journal} {Plasma Physics and Controlled Fusion}\
  }\textbf {\bibinfo {volume} {47}},\ \bibinfo {pages} {A231} (\bibinfo {year}
  {2005})}\BibitemShut {NoStop}%
\bibitem [{\citenamefont {Nieter}\ \emph {et~al.}(2009)\citenamefont {Nieter},
  \citenamefont {Cary}, \citenamefont {Werner}, \citenamefont {Smithe},\ and\
  \citenamefont {Stoltz}}]{nieter2009application}%
  \BibitemOpen
  \bibfield  {author} {\bibinfo {author} {\bibfnamefont {C.}~\bibnamefont
  {Nieter}}, \bibinfo {author} {\bibfnamefont {J.~R.}\ \bibnamefont {Cary}},
  \bibinfo {author} {\bibfnamefont {G.~R.}\ \bibnamefont {Werner}}, \bibinfo
  {author} {\bibfnamefont {D.~N.}\ \bibnamefont {Smithe}}, \ and\ \bibinfo
  {author} {\bibfnamefont {P.~H.}\ \bibnamefont {Stoltz}},\ }\bibfield  {title}
  {\enquote {\bibinfo {title} {Application of dey--mittra conformal boundary
  algorithm to 3d electromagnetic modeling},}\ }\href@noop {} {\bibfield
  {journal} {\bibinfo  {journal} {Journal of Computational Physics}\ }\textbf
  {\bibinfo {volume} {228}},\ \bibinfo {pages} {7902--7916} (\bibinfo {year}
  {2009})}\BibitemShut {NoStop}%
\bibitem [{\citenamefont {Squire}, \citenamefont {Qin},\ and\ \citenamefont
  {Tang}(2012)}]{squire2012geometric}%
  \BibitemOpen
  \bibfield  {author} {\bibinfo {author} {\bibfnamefont {J.}~\bibnamefont
  {Squire}}, \bibinfo {author} {\bibfnamefont {H.}~\bibnamefont {Qin}}, \ and\
  \bibinfo {author} {\bibfnamefont {W.~M.}\ \bibnamefont {Tang}},\ }\bibfield
  {title} {\enquote {\bibinfo {title} {Geometric integration of the
  vlasov-maxwell system with a variational particle-in-cell scheme},}\
  }\href@noop {} {\bibfield  {journal} {\bibinfo  {journal} {Physics of
  Plasmas}\ }\textbf {\bibinfo {volume} {19}},\ \bibinfo {pages} {084501}
  (\bibinfo {year} {2012})}\BibitemShut {NoStop}%
\bibitem [{\citenamefont {Monk}(2003)}]{monk2003finite}%
  \BibitemOpen
  \bibfield  {author} {\bibinfo {author} {\bibfnamefont {P.}~\bibnamefont
  {Monk}},\ }\href@noop {} {\emph {\bibinfo {title} {Finite element methods for
  Maxwell's equations}}}\ (\bibinfo  {publisher} {Oxford University Press},\
  \bibinfo {year} {2003})\BibitemShut {NoStop}%
\bibitem [{\citenamefont {Glasser}\ and\ \citenamefont
  {Qin}(2019)}]{glasser2019geometric}%
  \BibitemOpen
  \bibfield  {author} {\bibinfo {author} {\bibfnamefont {A.~S.}\ \bibnamefont
  {Glasser}}\ and\ \bibinfo {author} {\bibfnamefont {H.}~\bibnamefont {Qin}},\
  }\bibfield  {title} {\enquote {\bibinfo {title} {The geometric theory of
  charge conservation in particle-in-cell simulations},}\ }\href@noop {}
  {\bibfield  {journal} {\bibinfo  {journal} {arXiv preprint arXiv:1910.12395}\
  } (\bibinfo {year} {2019})}\BibitemShut {NoStop}%
\bibitem [{\citenamefont {Meierbachtol}\ \emph {et~al.}(2015)\citenamefont
  {Meierbachtol}, \citenamefont {Greenwood}, \citenamefont {Verboncoeur},\ and\
  \citenamefont {Shanker}}]{meierbachtol2015conformal}%
  \BibitemOpen
  \bibfield  {author} {\bibinfo {author} {\bibfnamefont {C.~S.}\ \bibnamefont
  {Meierbachtol}}, \bibinfo {author} {\bibfnamefont {A.~D.}\ \bibnamefont
  {Greenwood}}, \bibinfo {author} {\bibfnamefont {J.~P.}\ \bibnamefont
  {Verboncoeur}}, \ and\ \bibinfo {author} {\bibfnamefont {B.}~\bibnamefont
  {Shanker}},\ }\bibfield  {title} {\enquote {\bibinfo {title} {Conformal
  electromagnetic particle in cell: A review},}\ }\href@noop {} {\bibfield
  {journal} {\bibinfo  {journal} {IEEE Transactions on Plasma Science}\
  }\textbf {\bibinfo {volume} {43}},\ \bibinfo {pages} {3778--3793} (\bibinfo
  {year} {2015})}\BibitemShut {NoStop}%
\bibitem [{\citenamefont {Jin}(2015)}]{jin2015finite}%
  \BibitemOpen
  \bibfield  {author} {\bibinfo {author} {\bibfnamefont {J.-M.}\ \bibnamefont
  {Jin}},\ }\href@noop {} {\emph {\bibinfo {title} {The finite element method
  in electromagnetics}}}\ (\bibinfo  {publisher} {John Wiley \& Sons},\
  \bibinfo {year} {2015})\BibitemShut {NoStop}%
\bibitem [{\citenamefont {Pinto}\ \emph {et~al.}(2014)\citenamefont {Pinto},
  \citenamefont {Jund}, \citenamefont {Salmon},\ and\ \citenamefont
  {Sonnendr{\"u}cker}}]{pinto2014charge}%
  \BibitemOpen
  \bibfield  {author} {\bibinfo {author} {\bibfnamefont {M.~C.}\ \bibnamefont
  {Pinto}}, \bibinfo {author} {\bibfnamefont {S.}~\bibnamefont {Jund}},
  \bibinfo {author} {\bibfnamefont {S.}~\bibnamefont {Salmon}}, \ and\ \bibinfo
  {author} {\bibfnamefont {E.}~\bibnamefont {Sonnendr{\"u}cker}},\ }\bibfield
  {title} {\enquote {\bibinfo {title} {Charge-conserving fem--pic schemes on
  general grids},}\ }\href@noop {} {\bibfield  {journal} {\bibinfo  {journal}
  {Comptes Rendus Mecanique}\ }\textbf {\bibinfo {volume} {342}},\ \bibinfo
  {pages} {570--582} (\bibinfo {year} {2014})}\BibitemShut {NoStop}%
\bibitem [{\citenamefont {Moon}, \citenamefont {Teixeira},\ and\ \citenamefont
  {Omelchenko}(2015)}]{moon2014exact}%
  \BibitemOpen
  \bibfield  {author} {\bibinfo {author} {\bibfnamefont {H.}~\bibnamefont
  {Moon}}, \bibinfo {author} {\bibfnamefont {F.~L.}\ \bibnamefont {Teixeira}},
  \ and\ \bibinfo {author} {\bibfnamefont {Y.~A.}\ \bibnamefont {Omelchenko}},\
  }\bibfield  {title} {\enquote {\bibinfo {title} {Exact charge-conserving
  scatter--gather algorithm for particle-in-cell simulations on unstructured
  grids: A geometric perspective},}\ }\href@noop {} {\bibfield  {journal}
  {\bibinfo  {journal} {Computer Physics Communications}\ }\textbf {\bibinfo
  {volume} {194}},\ \bibinfo {pages} {43--53} (\bibinfo {year}
  {2015})}\BibitemShut {NoStop}%
\bibitem [{\citenamefont {Munz}\ \emph {et~al.}(2000)\citenamefont {Munz},
  \citenamefont {Omnes}, \citenamefont {Schneider}, \citenamefont
  {Sonnendr{\"u}cker},\ and\ \citenamefont {Voss}}]{munz2000divergence}%
  \BibitemOpen
  \bibfield  {author} {\bibinfo {author} {\bibfnamefont {C.-D.}\ \bibnamefont
  {Munz}}, \bibinfo {author} {\bibfnamefont {P.}~\bibnamefont {Omnes}},
  \bibinfo {author} {\bibfnamefont {R.}~\bibnamefont {Schneider}}, \bibinfo
  {author} {\bibfnamefont {E.}~\bibnamefont {Sonnendr{\"u}cker}}, \ and\
  \bibinfo {author} {\bibfnamefont {U.}~\bibnamefont {Voss}},\ }\bibfield
  {title} {\enquote {\bibinfo {title} {Divergence correction techniques for
  maxwell solvers based on a hyperbolic model},}\ }\href@noop {} {\bibfield
  {journal} {\bibinfo  {journal} {Journal of Computational Physics}\ }\textbf
  {\bibinfo {volume} {161}},\ \bibinfo {pages} {484--511} (\bibinfo {year}
  {2000})}\BibitemShut {NoStop}%
\bibitem [{\citenamefont {Crawford}\ \emph {et~al.}(2021)\citenamefont
  {Crawford}, \citenamefont {O'Connor}, \citenamefont {Luginsland},\ and\
  \citenamefont {Shanker}}]{crawford2021rubrics}%
  \BibitemOpen
  \bibfield  {author} {\bibinfo {author} {\bibfnamefont {Z.~D.}\ \bibnamefont
  {Crawford}}, \bibinfo {author} {\bibfnamefont {S.}~\bibnamefont {O'Connor}},
  \bibinfo {author} {\bibfnamefont {J.}~\bibnamefont {Luginsland}}, \ and\
  \bibinfo {author} {\bibfnamefont {B.}~\bibnamefont {Shanker}},\ }\bibfield
  {title} {\enquote {\bibinfo {title} {Rubrics for charge conserving current
  mapping in finite element particle in cell methods},}\ }\href@noop {}
  {\bibfield  {journal} {\bibinfo  {journal} {arXiv preprint arXiv:2101.12128}\
  } (\bibinfo {year} {2021})}\BibitemShut {NoStop}%
\bibitem [{\citenamefont {Chen}, \citenamefont {Chac{\'o}n},\ and\
  \citenamefont {Barnes}(2011)}]{chen2011energy}%
  \BibitemOpen
  \bibfield  {author} {\bibinfo {author} {\bibfnamefont {G.}~\bibnamefont
  {Chen}}, \bibinfo {author} {\bibfnamefont {L.}~\bibnamefont {Chac{\'o}n}}, \
  and\ \bibinfo {author} {\bibfnamefont {D.~C.}\ \bibnamefont {Barnes}},\
  }\bibfield  {title} {\enquote {\bibinfo {title} {An energy-and
  charge-conserving, implicit, electrostatic particle-in-cell algorithm},}\
  }\href@noop {} {\bibfield  {journal} {\bibinfo  {journal} {Journal of
  Computational Physics}\ }\textbf {\bibinfo {volume} {230}},\ \bibinfo {pages}
  {7018--7036} (\bibinfo {year} {2011})}\BibitemShut {NoStop}%
\bibitem [{\citenamefont {Crawford}\ \emph {et~al.}(2020)\citenamefont
  {Crawford}, \citenamefont {Li}, \citenamefont {Christlieb},\ and\
  \citenamefont {Shanker}}]{crawford2020unconditionally}%
  \BibitemOpen
  \bibfield  {author} {\bibinfo {author} {\bibfnamefont {Z.}~\bibnamefont
  {Crawford}}, \bibinfo {author} {\bibfnamefont {J.}~\bibnamefont {Li}},
  \bibinfo {author} {\bibfnamefont {A.}~\bibnamefont {Christlieb}}, \ and\
  \bibinfo {author} {\bibfnamefont {B.}~\bibnamefont {Shanker}},\ }\bibfield
  {title} {\enquote {\bibinfo {title} {Unconditionally stable time stepping
  method for mixed finite element maxwell solvers},}\ }\href@noop {} {\bibfield
   {journal} {\bibinfo  {journal} {Progress In Electromagnetics Research}\
  }\textbf {\bibinfo {volume} {103}},\ \bibinfo {pages} {17--30} (\bibinfo
  {year} {2020})}\BibitemShut {NoStop}%
\bibitem [{\citenamefont {Zienkiewicz}(1977)}]{zienkiewicz1977new}%
  \BibitemOpen
  \bibfield  {author} {\bibinfo {author} {\bibfnamefont {O.~C.}\ \bibnamefont
  {Zienkiewicz}},\ }\bibfield  {title} {\enquote {\bibinfo {title} {A new look
  at the newmark, houbolt and other time stepping formulas. a weighted residual
  approach},}\ }\href@noop {} {\bibfield  {journal} {\bibinfo  {journal}
  {Earthquake Engineering \& Structural Dynamics}\ }\textbf {\bibinfo {volume}
  {5}},\ \bibinfo {pages} {413--418} (\bibinfo {year} {1977})}\BibitemShut
  {NoStop}%
\bibitem [{\citenamefont {Boris}(1970)}]{boris1970relativistic}%
  \BibitemOpen
  \bibfield  {author} {\bibinfo {author} {\bibfnamefont {J.~P.}\ \bibnamefont
  {Boris}},\ }\bibfield  {title} {\enquote {\bibinfo {title} {Relativistic
  plasma simulation-optimization of a hybrid code},}\ }in\ \href@noop {} {\emph
  {\bibinfo {booktitle} {Proc. Fourth Conf. Num. Sim. Plasmas}}}\ (\bibinfo
  {year} {1970})\ pp.\ \bibinfo {pages} {3--67}\BibitemShut {NoStop}%
\bibitem [{\citenamefont {O'Connor}\ \emph {et~al.}(2021)\citenamefont
  {O'Connor}, \citenamefont {Crawford}, \citenamefont {Verboncoeur},
  \citenamefont {Lugisland},\ and\ \citenamefont {Shanker}}]{o2021set}%
  \BibitemOpen
  \bibfield  {author} {\bibinfo {author} {\bibfnamefont {S.}~\bibnamefont
  {O'Connor}}, \bibinfo {author} {\bibfnamefont {Z.}~\bibnamefont {Crawford}},
  \bibinfo {author} {\bibfnamefont {J.}~\bibnamefont {Verboncoeur}}, \bibinfo
  {author} {\bibfnamefont {J.}~\bibnamefont {Lugisland}}, \ and\ \bibinfo
  {author} {\bibfnamefont {B.}~\bibnamefont {Shanker}},\ }\bibfield  {title}
  {\enquote {\bibinfo {title} {A set of benchmark tests for validation of 3d
  particle in cell methods},}\ }\href@noop {} {\bibfield  {journal} {\bibinfo
  {journal} {arXiv preprint arXiv:2101.09299}\ } (\bibinfo {year}
  {2021})}\BibitemShut {NoStop}%
\bibitem [{\citenamefont {Kovalev}\ and\ \citenamefont
  {Bychenkov}(2003)}]{kovalev2003analytic}%
  \BibitemOpen
  \bibfield  {author} {\bibinfo {author} {\bibfnamefont {V.}~\bibnamefont
  {Kovalev}}\ and\ \bibinfo {author} {\bibfnamefont {V.~Y.}\ \bibnamefont
  {Bychenkov}},\ }\bibfield  {title} {\enquote {\bibinfo {title} {Analytic
  solutions to the vlasov equations for expanding plasmas},}\ }\href@noop {}
  {\bibfield  {journal} {\bibinfo  {journal} {Physical review letters}\
  }\textbf {\bibinfo {volume} {90}},\ \bibinfo {pages} {185004} (\bibinfo
  {year} {2003})}\BibitemShut {NoStop}%
\end{thebibliography}%

\end{document}